\documentclass[prb,twocolumn,floatfix,preprintnumbers,amsmath,amssymb,superscriptaddress]{revtex4}

\usepackage{graphicx}
\usepackage{dcolumn}
\usepackage{amsfonts}
\usepackage{bm}
\usepackage{amsmath}
\usepackage[symbol]{footmisc}
\def \NCTO{Na$_{2}$Co$_{2}$TeO$_{6}$}
\def \NNTO{Na$_{2}$Ni$_{2}$TeO$_{6}$}
\def \NxNTO{Na$_{2.4}$Ni$_{2}$TeO$_{6}$}

\begin{document}



\title{Static and dynamic magnetic properties of honeycomb lattice antiferromagnets Na$_{2}M_{2}$TeO$_{6}$, $M$ = Co and Ni
\footnote{This manuscript has been authored by UT-Battelle, LLC under Contract No. DE-AC05-00OR22725 with the U.S. Department of Energy.  The United States Government retains and the publisher, by accepting the article for publication, acknowledges that the United States Government retains a non-exclusive, paid-up, irrevocable, world-wide license to publish or reproduce the published form of this manuscript, or allow others to do so, for United States Government purposes.  The Department of Energy will provide public access to these results of federally sponsored research in accordance with the DOE Public Access Plan (http://energy.gov/downloads/doe-public-access-plan).}}

\author{Anjana M. Samarakoon}
\thanks{Current address: Materials Science Division, Argonne National Laboratory, Lemont, IL 60439, USA}
\affiliation{Neutron Scattering Division, Oak Ridge National Laboratory, Oak Ridge, TN 37831, USA}
\author{Qiang Chen}
\affiliation{Dep. Physics and Astronomy, University of Tennessee, Knoxville, TN 37996}
\author{Haidong Zhou}
\affiliation{Dep. Physics and Astronomy, University of Tennessee, Knoxville, TN 37996}
\author{V. Ovidiu Garlea}
\email[Corresponding author: ]{garleao@ornl.gov}
\affiliation{Neutron Scattering Division, Oak Ridge National Laboratory, Oak Ridge, TN 37831, USA}


\begin{abstract}

The magnetic structures and spin dynamics of \NCTO~and \NNTO~ are investigated by means of elastic and inelastic neutron scattering measurements and the results are discussed in the context of a generalized Kitaev-Heisenberg model on a honeycomb lattice with strong spin-orbit coupling. The large number of parameters involved in the Hamiltonian model is evaluated by using an iterative optimization algorithm capable of extracting model solutions and simultaneously estimating their uncertainty. The analyses indicate that both Co$^{2+}$ ($d^7$) and Ni$^{2+}$ ($d^8$) antiferromagnets realize bond-dependent anisotropic nearest-neighbor interactions, and support the theoretical predictions for the realization of Kitaev physics in 3$d$ electron systems with effective spins $S$=1/2 and $S$=1. By studying the Na-doped system \NxNTO, we show that the control of Na content can provide an effective route for fine tuning the magnetic lattice dimensionality, as well as to controlling the bond-dependent anisotropic interactions.

\end{abstract}


\maketitle
\section{Introduction}

The role of magnetic frustration arising from competing bond-dependent anisotropic interactions in the magnetic properties of honeycomb systems is the subject of intense research. Novel materials with strong spin-orbit coupling that can give rise to such interactions, called Kitaev-type interactions, are continually sought after. For an ideal Kitaev model, the spins fractionalize into Majorana fermions and form a topological quantum spin liquid (QSL).\cite{Kitaev} Yet most of the studied materials exhibit long-range magnetic orders at low temperatures and extended models accounting for competing anisotropic Kitaev and isotropic Heisenberg interactions have been employed.\cite{Kimchi,Winter,Maksimov,Khaliullin} A great deal of theoretical and experimental studies have been focused on spin–orbit-coupled 4$d$ and 5$d$ transition-metal-based Mott-insulating materials with honeycomb structure and effective spin $J_{eff}$=1/2.\cite{Jackeli,Rau,Takagi,Motome_rev} In the recent years, however, attempts have been made to extend the Kitaev model to 3$d$ transition metal-based materials. The cobaltates systems with Co$^{2+}$ in $d^7$ state possessing a pseudo-spin-1/2 were among the first candidates to be considered.\cite{Khaliullin,Liu,Sano,Motome} The presence of spin-active $e_g$ electrons in high-spin $d^7$ systems changes the balance between Kitaev and Heisenberg couplings. A proximity to charge-transfer insulating regime is expected to suppress the Heisenberg interactions and stabilize a QSL phase. It has been also argued that as one moves from $5d$ to $4d$ and to $3d$, magnetic $d$ orbitals become more localized, improving the condition to obtain the nearest-neighbor-only interaction model proposed by Kitaev.\cite{Khaliullin} Besides the vested interest in the $J_{eff}$ = 1/2 Kitaev spin liquids, recent theoretical studies have been devoted to the investigation of the Kitaev model with higher spins ($S$ = 1 or 3/2) as another possibility of realizing a QSL state.\cite{Stavropoulos,Dong,Lee} A microscopic mechanism for achieving high-spin Kitaev physics in systems with strong spin-orbit coupling in anions and strong Hund’s coupling in transition metal cations has been recently proposed.~\cite{Stavropoulos} The main candidate materials for this mechanism are based on honeycomb or triangular networks of $d^8$ transition metals, such as Ni$^{2+}$ with half filled $e_g$ orbitals.~\cite{Stavropoulos} The emergence of Kitaev interaction by means of spin–orbit coupling of the heavy ligands (e.g. iodine or tellurium) has also been also evidenced in several 2D materials containing Cr$^{3+}$ ions with $S$ = 3/2.~\cite{Xu,Bellaiche}

\begin{figure}[tbp]
\includegraphics[width=3.5in]{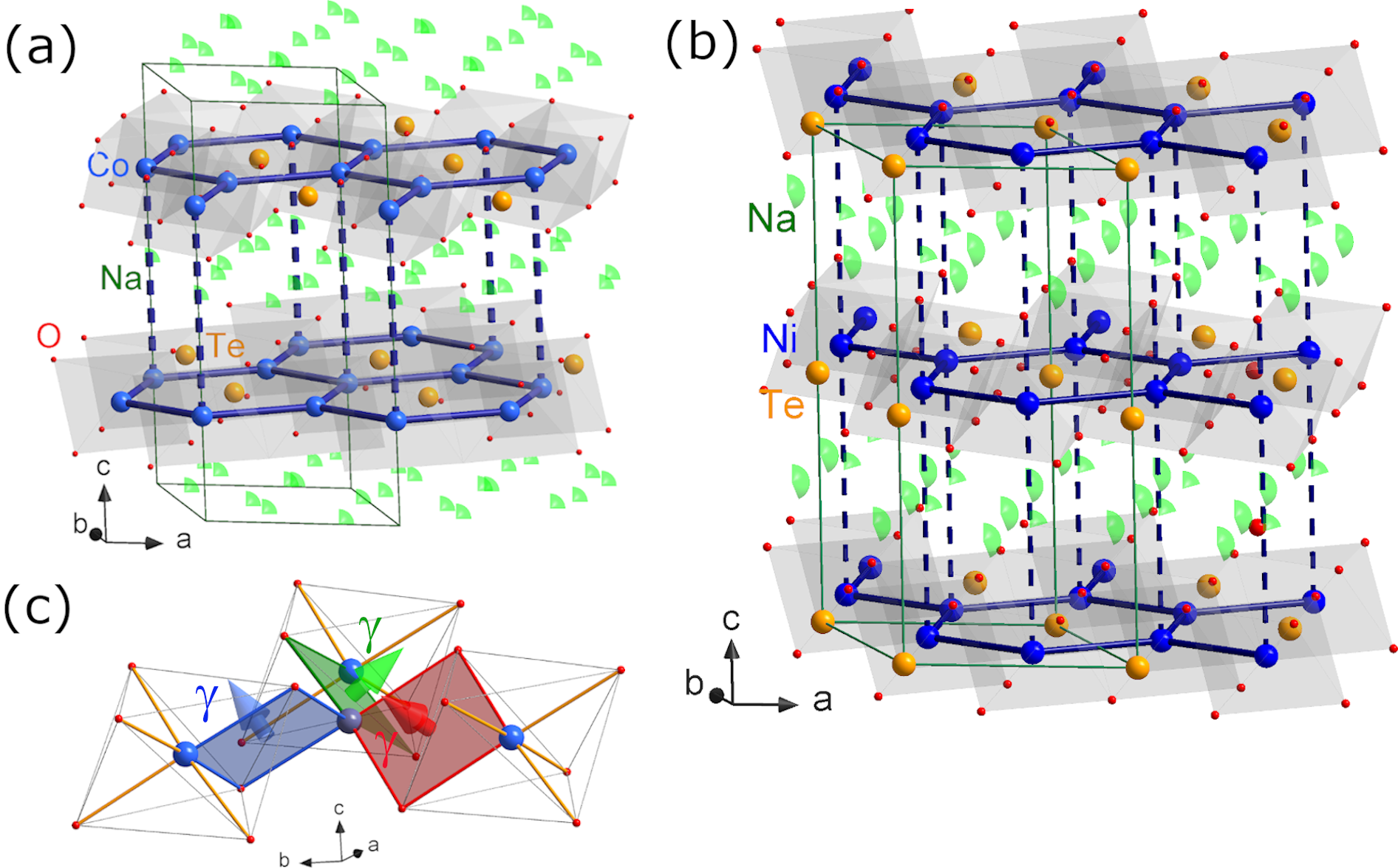}
\caption{\label{structure} Polyhedral view of (a) \NCTO~and (b) \NNTO~crystal structures consisting of two honeycomb layer separated by disordered Na atoms. In \NCTO~the successive Co honeycomb planes are shifted with respect to each other, while in \NNTO~the Ni atoms are stacked directly on top of one another. (c) View of the three adjacent nearest-neighbor (NN) bonds of the honeycomb lattice. The bond-dependent NN interactions are defined by mutually orthogonal Ising axes $\gamma$.}
\end{figure}

The honeycomb compounds with general formulas Na$_2M_2$TeO$_6$ and $A_3M_2X$O$_6$, where $M$ = Co$^{2+}$ or Ni$^{2+}$, $A$=Na, Li, Ag, and $X$= Sb or Bi, \cite{Viciu,Berthelot,Lefrancois,Bera,Yao,Karna,Sankar,Kurbakov,Korshunov,Kurbakov2,Seibel,Evstigneeva,Wong,Zvereva,Yan,Vivanco} are seen as some of the most promising $3d$ electron systems for the realization of the hybrid Kitaev-Heisenberg model. In these compounds, the honeycomb layers within the $ab$ plane are formed by edge-sharing $M$O$_6$ octahedra with (Te/Sb/Bi)O$_6$ at the center of the honeycomb lattice. The magnetic honeycomb layers are separated by the nonmagnetic layers of disordered $A$ = Na/Li/Ag atoms. The Na$_2M_2$TeO$_6$ compounds contain two honeycomb layers in an hexagonal unit cell, while $A_3M_2X$O$_6$ have a single honeycomb layer in a monoclinic lattice. Detailed structural studies revealed that the Co and Ni variants of Na$_2M_2$TeO$_6$ present different stacking arrangements of the honeycomb layers. The \NNTO~ crystal structure is defined by $P6_3/mcm$ space-group and consists of Ni honeycomb layers stacked directly on top of one another. For \NCTO,~ the structure is described by $P6_322$ space-group and the successive Co honeycomb planes are translated by [1/3, 2/3, 0] to bring Te atoms on top of one Co position. It must also be noted that the regular honeycomb networks are formed by a single crystallographic Ni site in \NNTO,~but by two symmetry-independent Co sites in \NCTO. The crystal structures of \NCTO~and \NNTO~are displayed in Fig.\ref{structure}. Despite the difference in stacking sequences and lattice symmetry all these compounds order magnetically at low temperatures in a zigzag-antiferromagnetic structure. The magnetic excitations in Na$_2$Co$_2$TeO$_6$ and Na$_2$Co$_2$TeO$_6$ compounds were investigated by using inelastic neutron scattering.\cite{sw_uk,sw_chi,sw_kor,Chen} Spin-orbit excitations observed in both compounds in the 20 - 28 meV energy range strongly support the premise that Co$^{2+}$ ions have a spin-orbital entangled $J_{eff}$=1/2 state. Those studies also demonstrated that a simple Heisenberg \textit{XXZ} model comprising first, second, and third nearest-neighbor couplings ($J_1$-$J_2$-$J_3$ model) fails to describe all features of the spectra. On the other hand, a Kitaev-Heisenberg Hamiltonian model with off-diagonal bond-directional interactions and long-range Heisenberg interactions gave a better match to the data. However, in absence of single crystal samples, the analyses of the powder averaged inelastic scattering produced some conflicting results regarding the nature of the Kitaev term, which was argued to be either ferromagnetic or antiferromagnetic.\cite{sw_uk,sw_chi,sw_kor}

In this paper, we reconsider the magnetic orders of both \NCTO~and \NNTO~and investigate their spin-wave excitations using powder inelastic neutron scattering. Modeling of the inelastic spectra is performed using an iterative optimization algorithm that allows exploring models that cover a broad parameter range and simultaneously estimating their uncertainty. The analyses revealed that the Kitaev-Heisenberg Hamiltonian is indeed better suited for describing the magnetic excitations in both compounds and that the possibility of an experimental realization of Kitaev physics in 3$d$ electron systems remains open. Constraints and limitations of the model are also discussed. We also investigate the static and dynamic spin properties of the Na-doped compound \NxNTO~and demonstrate that the control of Na content can be used to fine tune the dimensionality of the magnetic lattice as well as the bond-dependent anisotropic interaction.

\section{Experimental Details}
The powder samples used in this study were prepared by conventional solid-state reaction  in a similar manner as described in Refs.~\onlinecite{Viciu} and ~\onlinecite{Berthelot}. Samples were characterized by x-ray and magnetization measurements. The DC magnetic susceptibility data were obtained with a Quantum Design superconducting quantum interference device (SQUID) magnetometer through a zero field cooling process and with applied magnetic field of 1 kOe.

Neutron powder diffraction measurements were conducted using the HB2A powder diffractometer at the High Flux Isotope Reactor.~\cite{hb2a} Data were collected on approximately 5 g samples held in cylindrical vanadium containers that were placed in a top-loading closed cycle refrigerator (CCR). Measurements were performed at multiple temperatures in the range 4 - 100 K using $\lambda$ = 1.54~\AA~ and 2.41~\AA~ monochromatic beams, provided by a vertically focused Ge monochromator.

Inelastic neutron scattering (INS) experiments were conducted using the HYSPEC direct chopper spectrometer at the Spallation Neutron Source.~\cite{Hyspec} Measurements were carried out on powder samples held in aluminum containers with 1~cm diameter. Data presented in this study was collected using an incident neutron energy $E_i=15$~meV and Fermi chopper frequency of 240~Hz. The Co sample was cooled down to 1.6~K using an Orange cryostat, and the Ni-sample was cooled to 5 K using a CCR.

Refinements of the nuclear and magnetic structures were carried out using the FullPprof software.~\cite{FullProf} Magnetic structures models have been constructed using the magnetic symmetry tools available at the Bilbao Crystallographic Server while the magnetic space-groups are given in Belov-Neronova-Smirnova (BNS) notation.\cite{BCS} Spin-wave calculation were performed using the linear spin wave theory with the program SpinW.\cite{SpinW}

\section{Results and Discussion}
\subsection{Macroscopic properties and static magnetic order}

\begin{figure*}[tbp]
\includegraphics[width=6.5in]{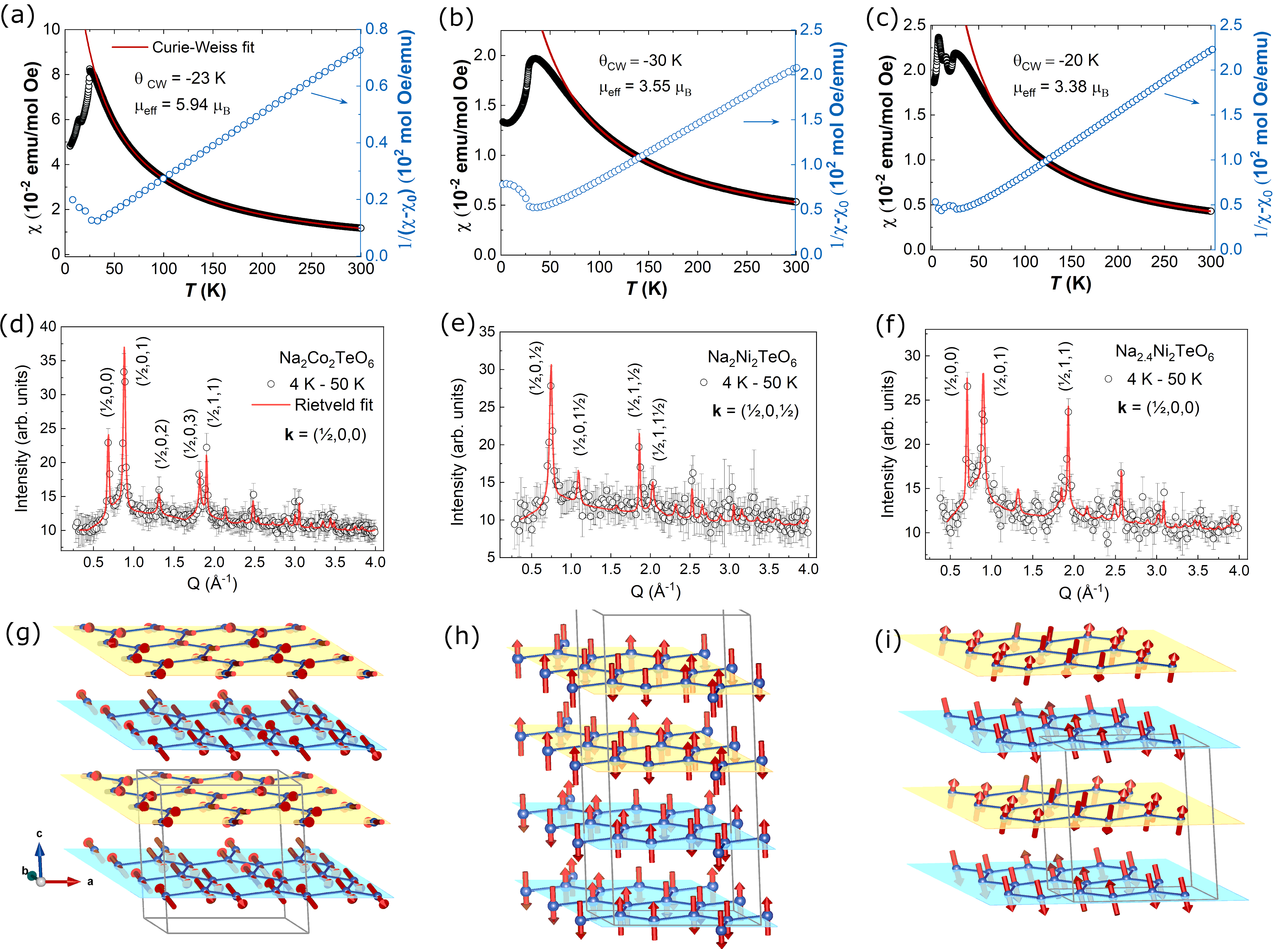}
\caption{\label{magstr}(a)(b)(c) DC magnetic susceptibility and its inverse measured under magnetic field of 1 kOe for \NCTO,~\NNTO~and \NxNTO,~respectively. (d)(e)(f) Magnetic scattering at 4 K obtained by subtracting the nuclear contribution measured at 50 K. Solid red lines represent the fit using the magnetic structure models described in the text. (g)(h)(i) Magnetic structures consisting of ferromagnetic zigzag chains that run along the $b$ direction (perpendicular to $\mathbf{k}$ vector). The magnetic moments are fully compensated within each honeycomb layer and are alternating their directions in successive layers. The staking sequence of adjacent honeycomb layers in \NCTO~(g) and \NxNTO~(i) is of \textit{A-B-A-B}-type, while in \NNTO~(h) is of a \textit{A-A-B-B}-type. The magnetic structure drawing were made using VESTA program.\cite{Vesta}}
\end{figure*}

The results of bulk magnetization measurements for \NCTO~ and \NNTO~ are shown in Figs.~\ref{magstr}(a) and (b). Magnetic ordering transitions are seen at low temperatures, near 25 K for \NCTO~  and at about 30 K for \NNTO,~ in good agreement with the earlier powder and single crystal studies.\cite{Viciu,Berthelot,Lefrancois,Bera,Yao,Karna,Sankar,Kurbakov} As previously observed, the Co compound displays an additional transition at approximately 15 K, which is thought to be associated with a spin reorientation. On the other hand, the AFM transition for \NNTO~appears to be smoother suggesting that a 2D short-range ordering precedes the 3D long-range order. Recent studies confirmed that a 2D order indeed emerges before the 3D order in both \NCTO~and \NNTO~compounds and persists to the lowest temperatures.\cite{Chen,Korshunov} Both \NCTO~and \NNTO~were shown to display some degree of structural disorder, and therefore, the short range-order is clearly due to a mixed contribution of weak exchange interactions and random interlayer bonds. Static susceptibility of the \NxNTO~displays common features to both Co and Ni parent compounds. A first broad transition appears at about 25 K and is followed by two additional spin reorientation transitions at 16 K and 6 K (see Fig.~\ref{magstr}(c)). The inverse susceptibility curves, shown as blue open symbols in Figs.~\ref{magstr}(a)(b)(c), are linear down to approximately 50 K and are fitted to a Curie-Weiss law with a temperature independent component $\chi_0$. For \NCTO~ the $\chi_0$ is found to be -2 x 10$^{-3}$ emu mol$^{-1}$ Oe$^{-1}$. This is very similar to that reported in Refs.\onlinecite{Viciu} and  \onlinecite{Lefrancois}, where it was assigned to the diamagnetic contribution coming from the sample and sample holder. For the Ni compounds we found $\chi_0$ = 5.2 x 10$^{-4}$ emu mol$^{-1}$ Oe$^{-1}$ for \NCTO~and -1.8 x 10$^{-4}$ emu mol$^{-1}$ Oe$^{-1}$ for \NxNTO.~Similar to our finding, Berthelot \textit{et al}~\cite{Berthelot} and Kurbakov \textit{et al}~\cite{Kurbakov} reported for \NNTO~a positive value of $\approx$ 1 x 10$^{-3}$ emu mol$^{-1}$ Oe$^{-1}$ that was attributed to the predominance of Ni$^{2+}$ Van Vleck paramagnetic contributions over diamagnetic contributions. The change in sign for the $\chi_0$ in \NxNTO~is likely due additional diamagnetism of the extra Na$^{+}$ ions. A similar sign change was found in the Zn$^{2+}$ doped samples Na$_2$Ni$_{2-x}$Zn$_x$TeO$_6$ ((0$< x <$1.5),~\cite{Berthelot} where $\chi_0$ changes from 1.1 x 10$^{-3}$ to -3.2 x 10$^{-4}$ emu mol$^{-1}$ Oe$^{-1}$. The effective magnetic moments inferred from the linear fits are $\mu_{eff}$ = 5.94 $\mu_B$/Co for \NCTO,~ and $\mu_{eff}$ = 3.55 $\mu_B$ and 3.38 $\mu_B$/Ni for the parent and off-stoichiometric Ni compound, respectively. The obtained Curie-Weiss temperature is $\Theta_{CW}$ = -23 K for the Co, and -30 K and -20 K for the Ni systems. The obtained values are very close to those reported in the previous studies. It is important to note that the effective moments for both Co and Ni compounds can only be explained by an additional orbital contribution to the spin moment. This contribution could be due to higher-order terms in the interplay between crystalline- field and spin–orbit couplings and covalency effects.

Neutron diffraction measurements were carried out at different temperatures above and below the ordering transitions. The structural parameters, including lattice constants, atomic positions and occupancies, as well as the thermal parameters were refined for all three samples using the 50 K data. Refinement results are summarised in the supplementary information.\cite{SM}~The analyses confirmed the structural models used in the previous studies, with $P6_322$ space group symmetry describing the \NCTO~crystal structure, and $P6_3/mcm$ for the \NNTO~and \NxNTO~structures. Special attention was given to Na site occupancies as it appeared to make an impact in the magnetic order of the Ni-based system. Due to the relatively weak magnetic scattering observed below the ordering transition, the analyses of the magnetic peaks were performed on the 4 K - 50 K subtracted data after the effect of magnetic form factor decay in paramagnetic contribution was properly considered. A systematic broadening for the $h, k, l\neq$0 magnetic reflections observed for all three samples, has been described using an anisotropic microstrain broadening model. The isolated magnetic scattering and the best fits are shown in Figs.~\ref{magstr}(d)(e)(f). Possible intermediate magnetic states closer to the N\'{e}el transition temperatures, indicated by susceptibility measurements, have not been explored due to insufficient statistics in the magnetic scattering. For the sake of clarity, the results of the magnetic refinements are discussed for each sample separately in the following sections.

\subsubsection{Magnetic order in \NCTO}

All the magnetic peaks of \NCTO~at 4 K were indexed with a propagation vector $\mathbf{k}$ = ($\frac{1}{2}$,~0,~0) (see Fig.~\ref{magstr}(d)). There are four possible maximal magnetic space groups that can describe the magnetic order from this $\mathbf{k}$-vector. The magnetic space group $P_C2_12_12_1$  (\#19.29) in a (2a,b,c) unit cell base gives the best fit to the data and produces the zigzag structure model that was previously proposed. In this model, the magnetic moments form ferromagnetic zigzag chains that run along the $b$ direction (perpendicular to $\mathbf{k}$), with the magnetic moments aligned in the $bc$ plane. The magnetic moments are fully compensated within each honeycomb layer and are also alternating their directions in successive layers. It is important to point out that the magnetic symmetry allows for an out of plane component $m_c$ (i.e. $m$ = (0, $m_b$, $m_c$)) that has been neglected in the previous reports. This component is particularly relevant in the context of the Kitaev-type anisotropic bond-directional couplings model proposed for this system to explain the spin-dynamics spectrum. An unconstrained refinement of both moment components for the two distinct Co positions (Co1 and Co2, defined in Table S1) was not possible, and the out-of-plane components ($m_c$) was constrained to be equal. The refined magnetic moment components are $m_b$=2.07(7)$~\mu_B$ for Co1, $m_b$=1.95(10)$~\mu_B$ for Co2, and $m_c$ = 0.5(2)$~\mu_B$. These yield a total static moment nearly identical (within the uncertainty range) for the two sites: 2.1(1)$~\mu_B$/Co1 and 2.0(1)$~\mu_B$/Co2. These values are slightly lower than reported previously ($m_{Co1}$ = 2.7$~\mu_B$, and $m_{Co2}$=2.45$~\mu_B$ at 1.8 K), where only the in plane components were considered. The magnetic structure used to fit our \NCTO~data is depicted in Fig.~\ref{magstr}(g). A tabulated description of the moments arrangement in the magnetic unit cell is given in the supplementary information material.\cite{SM}

One should also note that a 3-$\mathbf{k}$ (i.e. $\mathbf{k_1}$ = ($\frac{1}{2}$,~0,~0), $\mathbf{k_2}$ = (0,~$\frac{1}{2}$,~0), $\mathbf{k_3}$ = ($\frac{1}{2}$,~-$\frac{1}{2}$,~0)) magnetic structure model was recently suggested by Chen \textit{et. al}.\cite{Chen}~In that case the magnetic order will consist of an 120 degrees spin arrangement with only 3/4 of Co-atoms carrying an ordered moment in the $ab$ plan. We are not discussing that model here as our powder data cannot distinguish between a multi-$\mathbf{k}$ structure and multi-$\mathbf{k}$ domain contribution.

\subsubsection{Magnetic order in \NNTO}

The magnetic peaks of \NNTO~at 4 K, shown in Fig.~\ref{magstr}(e), were indexed with $\mathbf{k}$ = ($\frac{1}{2}$,~0,~$\frac{1}{2}$). This is a different propagation vector than the reported $\mathbf{k}$ = ($\frac{1}{2}$,~0,~0) in previous studies.\cite{Karna,Kurbakov} The main difference in the ordered state lies in the staking sequence of adjacent honeycomb layers that changes from a \textit{A-B-A-B} type to \textit{A-A-B-B}, where \textit{A} and \textit{B} display opposite moment directions. This sequence is surprising considering the magnetic isolation of honeycomb layers by semi-disordered Na layers and it suggests the existence of effective second-nearest-neighbor interlayer interactions that are competing with the nearest-neighbor interactions.\cite{Garlea} It is plausible that the out of the plane coupling is very sensitive to both Na amount and its distribution inside the Na monolayer. Sodium ions were previously reported to be distributed over multiple Wyckoff positions, but our refinements revealed only two positions being occupied and an occupancy very close to the stoichiometric value 2.0$\pm$0.02. We found that about 76\% ions partially occupy the Wyckoff 12j site and about 24\% the 4c site of $P6_3/mcm$. This is likely leading to reduction of the possible interlayer couplings and to a less magnetic disorder. A similar Na distribution was reported in Ref.~\onlinecite{Kurbakov}, but the overall sample composition was slightly off-stoichiometric ($\approx$ 2.13) and the magnetic ordering $\mathbf{k}$-vector was found to be ($\frac{1}{2}$,~0,~0).

There are four maximal magnetic space groups for the parent space group $P6_3/mcm$ and the propagation vector $\mathbf{k}$ = ($\frac{1}{2}$,~0,~$\frac{1}{2}$). The best fitting model is given by the $I_amm2$ (\#44.234) magnetic space group in the unit cell (2a, b, 2c). All Ni atoms and the corresponding ordered moments are described by a single Wyckoff site. The magnetic symmetry allows for ordered components along all crystallographic axes but the refinements show that the moments are aligned parallel to the $c$-axis. Similar to the Co-compound the moments are arranged in a zigzag structure with ferromagnetic chains running along the $b$ direction. The refined value of the static moment is 1.55(6)~$\mu_B$/Ni, with an in-plane component evaluated to be of less than 0.05~$\mu_B$. The magnitude of the moment is smaller than the theoretically expected value for Ni$^{2+}$ with $S$ = 1. A graphical representation of \NNTO~magnetic structure is shown in Fig.~\ref{magstr}(h) and a detailed information on the spin arrangement is given in Table S5.

\subsubsection{Magnetic order in \NxNTO}

Rietveld refinements of \NxNTO~crystal structure revealed that the excess Na occupies an additional Wyckoff position, 2a (0,~0,~1/4), while the occupancies for the other two positions remained nearly unchanged. The additional disorder in the Na layer leads to a change in the magnetic lattice, with the adjacent magnetic honeycomb layers following a \textit{A-B-A-B} type of stacking. As presented in Fig.~\ref{magstr}(f), the magnetic peaks are described by the wave-vector $\mathbf{k}$ = ($\frac{1}{2}$,~0,~0). We determined that this system orders in the same zigzag-type magnetic structure, and that the structure is described by the magnetic space group $P_A nma$ (\#62.453) on the base of (2a, b, c) lattice setting. However, in contrast to the \NNTO,~the ordered moment exhibits a canting away from the $c$-axis by approximately 30 degrees. The magnetic symmetry constrains the moments to lie in the $bc$ plane and the refined components are: $m_b$=0.7(1)$~\mu_B$ and $m_c$=1.30(5)$~\mu_B$. The total magnitude of the static moment is 1.5(1)$~\mu_B$/Ni. The magnetic structure of \NNTO~ is displayed in Fig.~\ref{magstr}(i). Based on the magnetization data one could expect that \NxNTO~features spin reorientations with temperature-dependent canting, similar to that seen in the related monoclinic compound Li$_3$Ni$_2$SbO$_6$.\cite{Kurbakov2}

\subsection{Neutron inelastic scattering}

\begin{figure*}[tbp]
\includegraphics[width=7.0in]{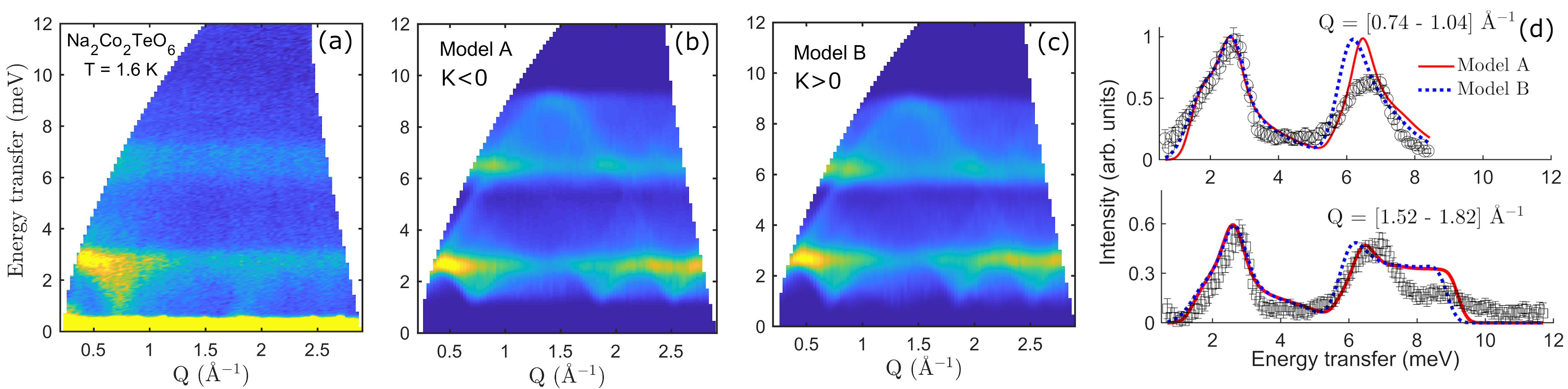}
\caption{\label{insCo} (a) Powder inelastic neutron spectrum of \NCTO~at T = 1.6 K measured on HYSPEC using E$_i$=15 meV. (b)(c) Calculated $S(Q,\omega)$ using the Kitaev-Heisenberg Hamiltonian corresponding to two optimized solutions with $K < 0$ (Model A) and $K > 0$ (Model B), respectively. (d) Comparison of the two selected models with the experimental data through cuts along energy transfer for two Q-integrated regions around 0.9 \AA$^{-1}$ and 1.67 \AA$^{-1}$. }
\end{figure*}

\begin{figure*}[tp]
\includegraphics[width=7.0in]{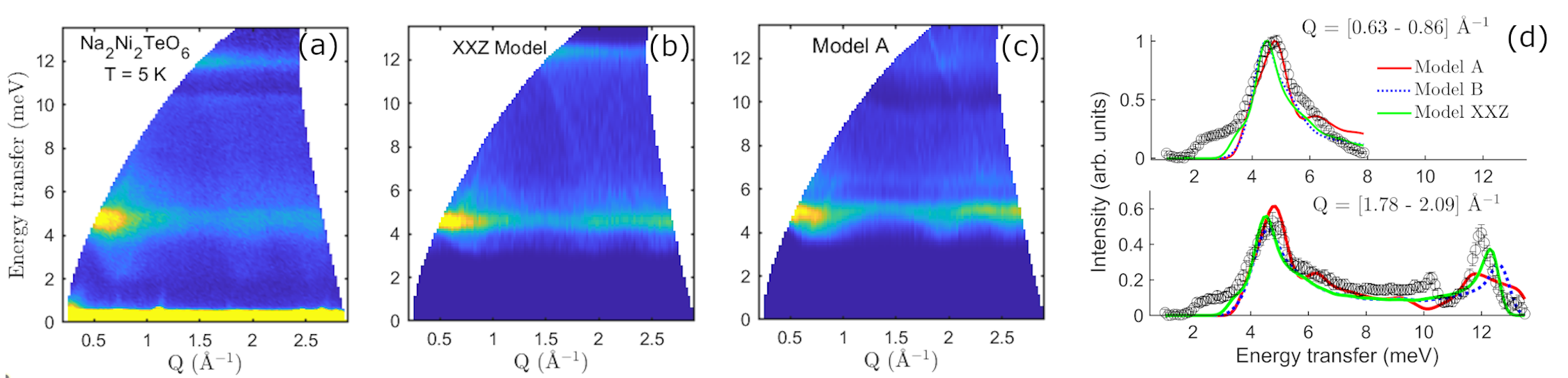}
\caption{\label{insNi}(a) Inelastic spectrum of \NNTO~measured at HYSPEC using E$_i$=15 meV. (b)(c) Calculated powder averaged spin-wave spectra using Heisenberg \textit{XXZ} and Kitaev- Hamiltonian models (d) Comparison of the \textit{XXZ} and \textit{K-H} models with the experimental data through cuts along energy transfer for two $Q$-integrated regions around 0.75 \AA$^{-1}$ and 1.9 \AA$^{-1}$. }
\end{figure*}

The inelastic neutron spectra of all three studied samples present two main modes: a gapped dispersive mode at low energies and a second flat mode at slightly higher energies. The contour maps of inelastic neutron scattering intensity in momentum-energy ($Q$-$E$) space measured using the incident neutron energy $E_i=15$~meV are shown in Figs.~\ref{insCo}(a),~\ref{insNi}(a),~\ref{insNix}(a). We modeled the data using a generalized Kitaev-Heisenberg (\textit{K-H}) Hamiltonian that accounts for bond-dependent anisotropic-exchange interactions, similar to that discussed in previous inelastic studies of \NCTO:~\cite{sw_uk,sw_chi,sw_kor}

\begin{equation}
\label{eq_1}
\begin{aligned}
\begin{split}
\mathcal{H}_{K-H}=& \sum_{{\left\langle{}i,j\right\rangle{}}_{r={1,2,3,4}}}  J_{r}S_i S_j
+\sum_{{\left\langle{}i,j\right\rangle{}}_1\in{\left\{\alpha,\beta,\gamma\right\}}}\left[ K S_i^{\gamma}\cdot S_j^{\gamma}\right.+ \\
&+ {\Gamma}\left(S_i^{\alpha} S_j^{\beta}+S_i^{\beta} S_j^{\alpha}\right)+ \\
& + \left.{{\Gamma}^{'}}\left(S_i^{\alpha} S_j^{\gamma}+S_i^{\gamma} S_j^{\alpha}+S_i^{\beta} S_j^{\gamma}+S_i^{\gamma} S_j^{\beta{}}\right)\right]+ \\
& +D\sum_{i}{\left(S_i\cdot \tilde{n_i}\right)}^2
\end{split}
\end{aligned}
\end{equation}

The bond notation ${\left\langle{}i,j\right\rangle{}}_r$ indicates that the corresponding sum runs over pairs of $r^{th}$ nearest neighbor (NN) bonds, including the first, second, and third in-plane NN couplings ($J_1$, $J_2$, $J_3$) and an inter-layer coupling ($J_4\equiv J_c$). There are three types of first NN bonds and the notation $\left\{\alpha,\beta,\gamma\right\}$ indicates that the sum runs over each of the three orthogonal bonds. $K$ represents the Kitaev interaction, and $\Gamma$ and $\Gamma^{\prime}$ are bond-dependent off-diagonal exchange interaction terms. Only the first NN exchange tensor is defined as anisotropic. The $D$ and $\widetilde{n_i}$ denotes the single-ion anisotropy and its direction. The single-ion anisotropy (SIA) term has only been used for the special case of $K\rightarrow0$ and $\Gamma=\Gamma^\prime$, when (\textit{K-H}) model is reduced to a \textit{XXZ}-type Hamiltonian.

\begin{figure}[btp]
\includegraphics[width=3.45in]{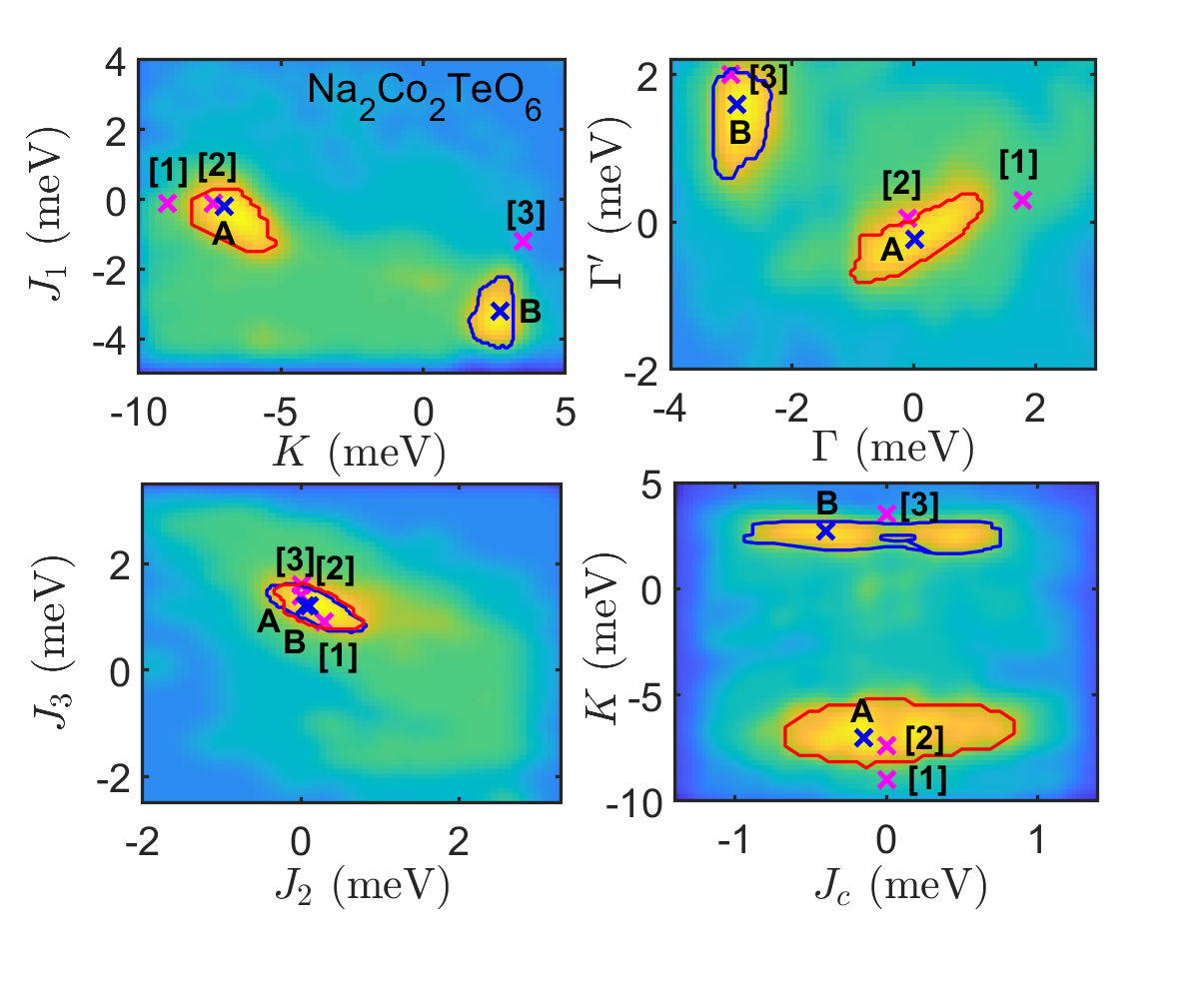}
\caption{\label{modelCo} The manifold of possible Kitaev-Heisenberg Hamiltonian solution for \NCTO.~Color maps represent the projected $\hat{\chi}_{INS}^2$ on 2D slices of parameter space in logarithmic scale. The region corresponding to FM Kitaev solution ($K<$0) is indicated by solid red line, while the AFM Kitaev solution ($K>$0) is depicted by a blue line. Two representative solutions of the two regions are indicated inside the regions with labels ``A'' and ``B''. The parameters reported in previous INS studies are also marked as [1],\cite{sw_uk}~,[2] and [3].~\cite{sw_kor}}
\end{figure}

Given the challenge of dealing with a high-dimensional (d$\leq$7) Hamiltonian space ($\mathcal{H}_{K-H}$ includes up to seven independent parameters: $J_1$,~$J_2$,~$J_3$,~$J_c$,~$\Gamma,~\Gamma^\prime$, and $K$ or $D$) and considering the significant information loss in the powder averaged INS data, it has been important to implement an optimization protocol for simultaneously extracting model solutions and estimating their uncertainty. To quantify the uncertainty of a proposed model solution, we applied the iterative optimization procedure explained in Ref.~\onlinecite{machine}. We used SpinW package to calculate the powder inelastic neutron scattering cross-section for a given Hamiltonian parameter set, based on linear spin-wave theory (LST). Calculations were preceded by a magnetic structure optimization starting from the refined structures discussed in the previous section. The cost function for the optimization process was: $\chi_{INS}^2$=$\sum_{\omega}\sum_{Q}{m\left(Q,\omega\right)\left(I_{exp.}\left(Q,\ \omega\right)-I_{cal.}\left(Q,\omega\right)\right)^2}$, with $m\left(Q,\omega\right)$ representing a step function to mask pixels either contaminated by direct beam or out of detector coverage. For each iteration we used random samples over the whole Hamiltonian space to build a low-cost estimator of $\chi_{INS}^2$, $\hat{\chi}_{INS}^2$. We then used $\hat{\chi}_{INS}^2$ to evaluate the next set of parameters uniformly distributed over the Hamiltonian space and subjected to the constraint $\hat{\chi}_{INS}^2<c$. The cutoff $c$ was lowered after each iteration. The last iteration was attained for a final value, $c_{final}$, for which the calculated intensity agreed with the INS data within the experimental uncertainty. The $c_{final}$ values were 0.5754 and 0.4206 for the Co and Ni data sets, respectively, corresponding to a 2.5\% error margin of whole parameter space. Additional details of the model description and fitting optimization process is given in the supplementary information.\cite{SM}~ In the following we detail the results obtained for each of the three studied compounds.

\subsubsection{Spin-wave excitations in \NCTO}

\begin{table}[btp]
\caption{\label{table1} Parameters of the generalized Kitaev-Heisenberg model used in this or previous studies to describe the \NCTO~spin-wave spectrum. The parameters values are given in meV. The labels used in the table correspond to those shown in the contour plots in Fig.~\ref{modelCo}.}
\begin{ruledtabular}
\begin{tabular}{ccccccccc}
\vspace{.1in}
Label & $K$ & $\Gamma$ & $\Gamma^{\prime}$ & $J_1$ &$J_2$ & $J_3$ & $J_c$ & Reference \\[3pt]
\hline \\[2pt]
1 &-9  & 1.8 & 0.3  &-0.1 &0.3 &0.9 & 0 &~\onlinecite{sw_uk}\\[3pt]
2 &-7.4& -0.1& 0.05&-0.1 &0   & 1.4 & 0&~\onlinecite{sw_kor}\\[3pt]
3 & 3.5& -3  & 2   &-1.2 &0   &1.6  & 0&~\onlinecite{sw_kor}\\[3pt]
A   &-7  & 0.02& -0.23&-0.2 &0.05&1.2 &-0.15& \\[3pt]
B   & 2.7& -2.9& 1.6  &-3.2 &0.1 &1.2 & -0.4& \\[3pt]
\end{tabular}
\end{ruledtabular}
\end{table}

The powder inelastic neutron spectrum of \NCTO~is shown in Fig.~\ref{insCo}(a). To explain the spin dynamics in this system both \textit{XXZ} and \textit{K-H} Hamiltonians have been considered.~\cite{sw_uk,sw_chi,sw_kor} Both models are capable of describing the low-energy dispersive mode extending to approximately 3 meV energy transfer. Previous inelastic studies have come to a consensus that the \textit{XXZ}-Heisenberg model fails in reproducing the correct bandwidth and $Q$-dependence of the higher energy mode located in the 6 - 8 meV energy range, while the \textit{K-H} Hamiltonian model appears more promising in that respect. Consequently, we focus only on the \textit{K-H} model in this work. As discussed above, the large number of parameters involved in the Kitaev model makes it  very challenging finding a unique solution, especially when dealing with powder-averaged data. Two of the previous INS studies~\cite{sw_uk,sw_chi} reported a ferromagnetic (FM) Kitaev coupling ($K <$ 0), while a third study suggested that the Kitaev coupling is antiferromagnetic ($K >$ 0).~\cite{sw_kor} In the latter study, an overestimate of the intensity of the high-energy mode was explained by an unaccounted damping effect originating from a two-magnon scattering process. In the same study, it has also been argued that only the AFM Kitaev model can stabilize the zigzag magnetic structure with moments aligned orthogonal to the $\mathbf{k}$-vector. The parameters determined in previous studies are summarized in Table~\ref{table1}.

To further improve the understanding of the magnetic interactions in~\NCTO,~we conducted a multi-dimensional parameter-space search described above on the suggested \textit{K-H} Hamiltonian. The spin-wave model assumed a magnetic form factor corresponding to Co$^{2+}$ magnetic ions and an effective spin $J_{eff}$ = 1/2. The manifold of possible parameter solutions is indicated by the contour plots in Fig.~\ref{modelCo}. We found that the Kitaev Hamiltonian can indeed yield solutions with different signs for the Kitaev parameter. Two equally-good solutions for the $K<$0 (FM) and $K>$0 (AFM) regions can be selected. The Hamiltonian parameters of the two representative solutions are marked in Fig.~\ref{modelCo} as ``A''  for $K<$0 and ``B'' for $K>$0. The parameter reported in the earlier studies are also indicated in the contour plots as [1], [2] and [3]. As visible in the figure, the previously reported values for the second and third NN exchange interactions ($J_2$, $J_3$) are in good agreement with the optimised parameter space regions obtained from our analysis. However, the relative values of $J_1$ and the Kitaev term ($K$), as well as the off-diagonal exchange interactions ($\Gamma$, ~$\Gamma^{\prime}$) are falling outside the optimal $\hat{\chi}_{INS}^2$ zone. One can also note that the interlayer coupling ($J_c$) is found in the case of $K<$0 to be distributed over a relatively broad range centered near zero value, whereas for the $K>$0 solution the optimal $J_c$ tends to nucleate away from zero. The actual values of Hamiltonian parameters corresponding to the two selected solutions are tabulated in Table~\ref{table1}, and the calculated powder average spectra are presented in Figs.~\ref{insCo}(b)(c). The spectra were convoluted the instrumental energy resolution described as a Gaussian function. Figure~\ref{insCo}(d) shows a comparison of the two models through two constant-$Q$ cuts superimposed to the experimental data. Interestingly, the calculated spectra for the two models are almost indistinguishable. Both models reproduces most of characteristic features measured experimentally, but they are deficient in describing the intensity distribution in high energy mode. It is important to point out that both solutions (FM and AFM $K$) were found to stabilize magnetic structures that are consistent with the diffraction results. Optimized magnetic structures are shown in supplementary information.\cite{SM}~To summarize, our results support the realization of bond-dependent anisotropic nearest-neighbor interactions in \NCTO,~ but also indicate that is impossible to select a unique model using the powder averaged INS data.

\begin{figure}[tbp]
\includegraphics[width=3.3in]{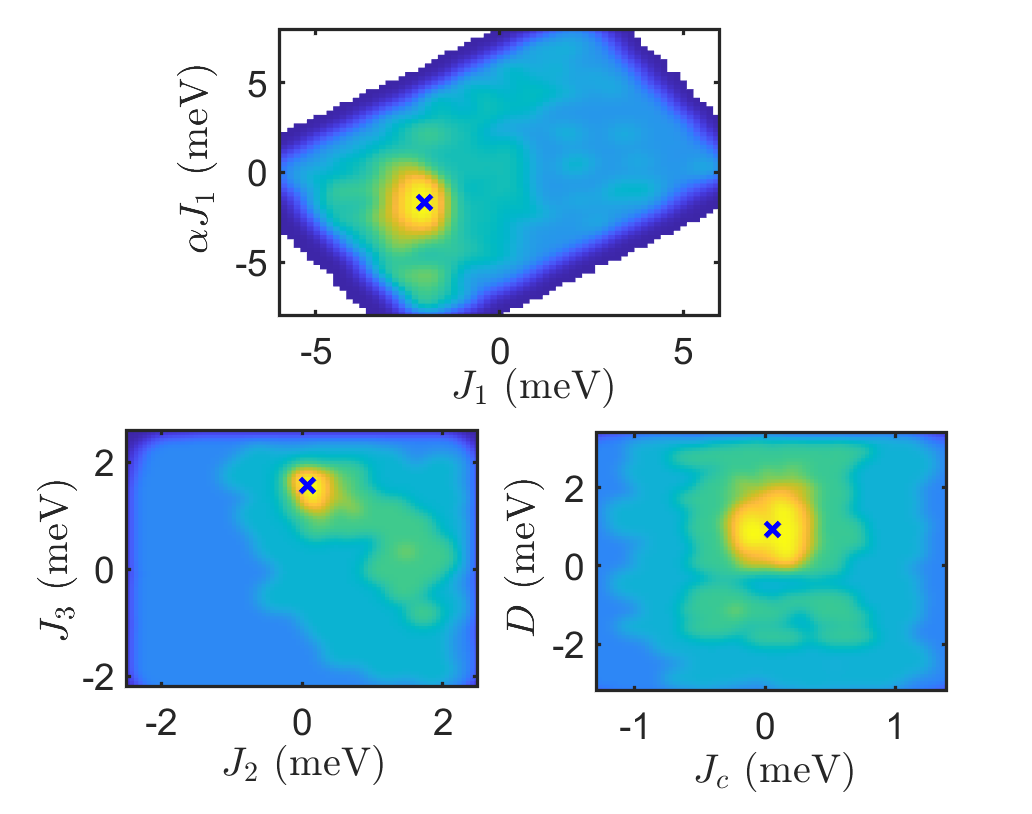}
\caption{\label{XXZmodelNi} Contour plots of projected $\hat{\chi}_{INS}^2$, showing the possible solutions for \NNTO~spin-wave spectrum using the \textit{XXZ} Hamiltonian model. The ``x'' symbol marks the parameters used for the $S(Q,\omega)$ simulation in Fig.~\ref{insNi}(b).  As described in the text, this model describes the main features of the magnetic excitations but fails to explain the gap opening in the high energy part of the spectrum.}
\end{figure}

\begin{figure}[tbp]
\includegraphics[width=3.45in]{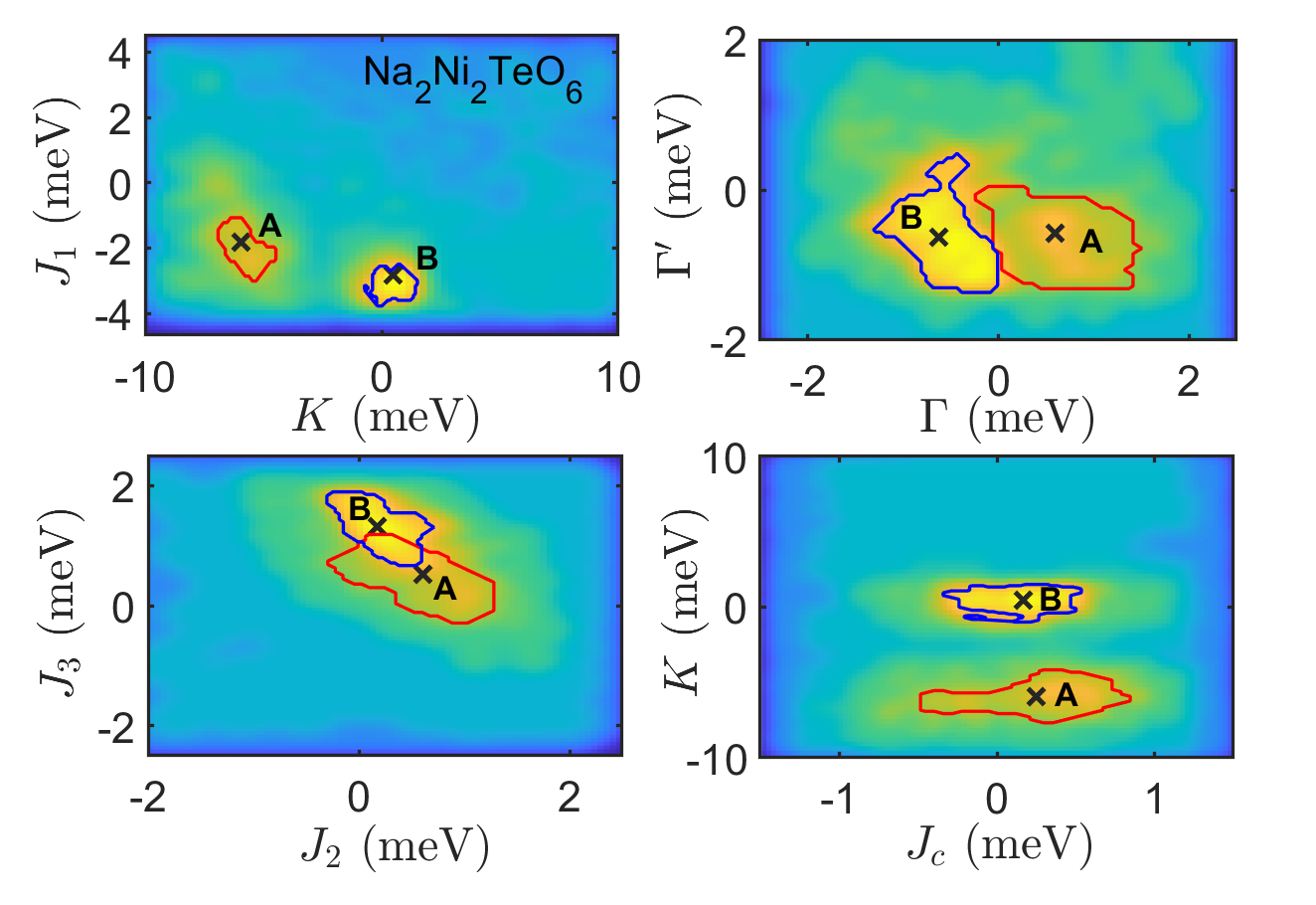}
\caption{\label{KmodelNi} The manifold of possible \textit{K-H} model solutions describing the magnetic excitations in \NNTO.~Two optimal regions have been identified: one corresponding to a FM $K$ solutions ($K<$0) is indicated by the red line, and a second region with $K\approx 0$ and $\Gamma \approx \Gamma^{\prime}$, indicated by the blue line. The second solution is nearly equivalent to the \textit{XXZ} model, but without a single-ion anisotropy term. For each regions a representative solution has been selected, labeled as ``A'' and ``B''.}
\end{figure}

\subsubsection{\NNTO}

The inelastic spectrum measured on \NNTO~powder at $T$ = 5~K is displayed in Fig.~\ref{insNi}(a). The spectrum consists of a gapped mode that extends to approximately 4.7 meV, and a second highly dispersive mode that reaches a maximum energy of about 12 meV. An interesting feature appears at about 11 meV energy transfer where an abrupt drop in the intensity occurs, appearing as a gap opening in the spectrum. A first attempt of describing the spin-wave spectrum was made using a \textit{XXZ}-Heisenberg Hamiltonian with a SIA term that forces the moment direction along the $c$-axis, as determined by diffraction. This model was obtained by imposing the constraints $K$=0 and $\Gamma=\Gamma^\prime$, in the generalized hamiltonian given in equation ~\ref{eq_1}. The adapted \textit{XXZ} Hamiltonian can be expressed as:

\begin{equation}
\label{eq_2}
\begin{aligned}
\begin{split}
\mathcal{H}_{XXZ}=& J_{1}\sum_{{\left\langle{}i,j\right\rangle{}}_1}\left(S_i^x S_j^x + S_i^y S_j^y +\alpha S_i^z S_j^z\right)+ \\
&+ \sum_{{\left\langle{}i,j\right\rangle{}}_{r=2,3,4}} J_{r}S_i S_j + D\sum{\left(S_i\cdot \tilde{c_i}\right)}^2
\end{split}
\end{aligned}
\end{equation}

where $\alpha$ is the spin anisotropy parameter, $J_4$ ($\equiv J_c$) is the nearest-neighbors (NN) interlayer interaction, and $D$ is the easy axis anisotropy along $c$-axis. Note that only the first NN interaction $J_1$ is considered anisotropic. One should also mention that we ignored the next-nearest-neighbor interlayer interactions that would be needed to stabilized a \textit{A-A-B-B} staking sequence since their contribution would likely be too small to be evaluated using the available data. The spin-wave calculations assumed a magnetic form factor corresponding to Ni$^{2+}$ and a spin value $S$ = 1. The distribution of best fitting parameters obtained for the \textit{XXZ}-model is shown in the contour plots in Fig.~\ref{XXZmodelNi}. A possible solution inside the optimized region is represented by the following parameters: $J_1$ = -2.051 meV, $\alpha$ = 0.8, $J_2$ = 0.081 meV, $J_3$ = 1.56 meV, $J_c$ = 0.055 meV and $D$ = -0.93 meV. This solution is consistent with what is expected for a zigzag spin structure, requiring ferromagnetic first NN and antiferromagnetic third-NN interactions. The corresponding calculated powder averaged spectrum for this solution is shown in Fig.~\ref{insNi}(b). As visible in the figure, the model gives a satisfactory description of the main features of the magnetic excitations, but fails in reproducing the split of the spectrum seen at about 11 meV.

In order to capture the gap opening, we next considered the \textit{K-H} Hamiltonian. Note that the spin gaps in the spectrum could also be accounted for using Dzyaloshinskii- Moriya (DM) interaction that occurs on the bonds without inversion symmetry.\cite{CrI3} However, DM interaction is absent in our Ni-systems due to the presence of an inversion center between first and third nearest-neighbor Ni ions. The manifold of acceptable solutions for the Kitaev model is represented in Fig.~\ref{KmodelNi} as color plots of the projected $\hat{\chi}_{INS}^2$ into two-dimensional slices of the parameter space. Similar to the Co-system the manifold consists of two distinctive regions, except that only one of them is localized at $K \neq 0$. A second region with $K\approx 0$ and $\Gamma \approx \Gamma^{\prime}$  appears to nearly coincide with the \textit{XXZ} model. The two optimal regions are marked in Fig.~\ref{KmodelNi} by different colors: blue contour line for $K\approx 0$ and red line for $K \neq 0$. Two representative solutions ``A'' and ``B'' were selected, with the corresponding parameters values shown in Table~\ref{table2}.

\begin{table}[btp]
\caption{\label{table2} Parameters of the \textit{XXY} and \textit{K-H} models used to describe the \NNTO~spin-wave spectrum. The values are given in meV units.}
\begin{ruledtabular}
\begin{tabular}{ccccccccc}
\vspace{.1in}
Label & $K$ & $\Gamma$ & $\Gamma^{\prime}$ & $J1$ &$J2$ & $J3$ & $Jz$ & $D$  \\[3pt]
\hline \\[1pt]
\textit{XXZ} &0   & 0.132 & 0.132 & -1.92 &0.081  & 1.56 & 0.055 & -0.93 \\[3pt]
Model A    &-5.95& 0.59  & -0.58  &-1.83  &0.604  & 0.524 & 0.25  & 0    \\[3pt]
Model B    & 0.49& -0.63 & -0.65  &-2.856 &0.172  & 1.316  & 0.166  & 0    \\[3pt]
\end{tabular}
\end{ruledtabular}
\end{table}

The optimal solution ``A'' is located at ferromagnetic side of $K$ and $\Gamma$, and antiferromagnetic side of $\Gamma^{\prime}$.  The $\Gamma$ and $\Gamma^{\prime}$ are comparable in magnitude but have opposite signs. The spin-wave spectrum obtained from this model is shown in Fig.~\ref{insNi}(c). A direct comparison of the best fitting  models is shown in Figure~\ref{insNi}(d), by superimposed constant-$Q$ cuts through $S(Q,\omega)$ integrated over the ranges [0.63 - 0.86] and [1.8 - 2.1] \AA$^{-1}$. Similar to \textit{XXZ}-model, the optimal solution ``B'' captures the main features in $S(Q,\omega)$, but fails to predict the energy gap at ~11 meV. In contrast, the solution ``A''  is successful in providing a qualitative explanation for the gap opening at high energies.

The possibility of realization of spin-1 Kitaev spin model in layered transition metal oxides has been recently discussed by Stavropoulos \textit{et. al}.~\cite{Stavropoulos}. The authors identified the honeycomb transitional metal oxide compounds $A_3$Ni$_2X$O$_6$ ($A$ = Li, Na, $X$ = Bi, Sb), which are isostructural with \NNTO,~as potential candidates for such Kitaev model. The bond-dependent interactions are generated via superexchange between two Ni$^{2+}$ cations with $e_g$ orbitals mediated by anion $p$ orbital electrons with a strong spin-orbit coupling induced by the proximity to the heavy Te or Sb atoms.

The optimized magnetic structure for the \textit{K-H} model ``A'' is of a zigzag type with magnetic moments canted away from the $c$-axis (see Fig S3). Stabilizing the magnetic order with spins parallel to $c$ axis, as inferred from the diffraction data, would require to include an easy-axis anisotropy. An interplay between Kitaev interaction and single-ion anisotropy cannot be excluded since such a mechanism was previously proposed to naturally explain the different magnetic behaviors on CrI$_3$ and CrGeTe$_3$.\cite{Xu} First-principles calculation carried out in those systems indicate that the Iodide or Tellurium ligands could enhance the spin-orbit coupling to produce not only Kitaev interactions but also strong single-ion anisotropies. Unfortunately, the interplay between Kitaev interaction and single-ion anisotropy cannot be investigated in our system using our powder averaged INS data due to the strong correlation between the Kitaev (K) and easy-axis anisotropy (D) parameters. Despite the \textit{K-H} model's limitation in reproducing the exact magnetic order, it provides a promising starting point for more sophisticated models that will need to be applied when single crystal INS data becomes available.

\begin{figure}[btp]
\includegraphics[width=3.35in]{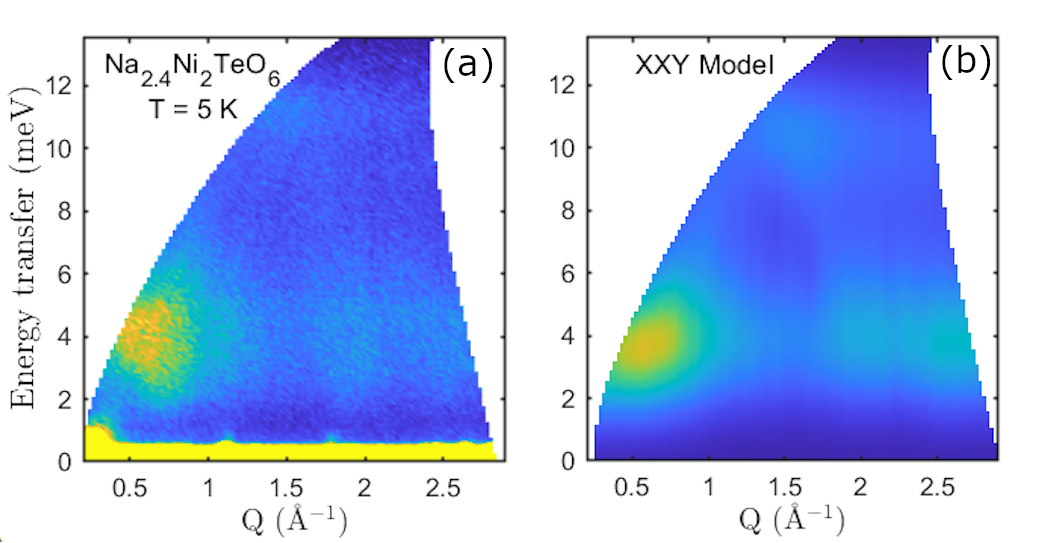}
\caption{\label{insNix}(a) Experimental inelastic spectrum of \NxNTO~measured at 5 K. (b) Calculated powder averaged spin-wave spectrum using a $S$=1 $XXZ$ Heisenberg model.}
\end{figure}

\subsubsection{\NxNTO}

The experimental INS data collected on the Na-doped sample \NxNTO~at $T$=5 K is presented in Fig.~\ref{insNix}(a). The magnetic excitations appear to be much broader than those observed on \NNTO,~in both energy transfer and $Q$ dimensions. The center of mass of the spectrum is slightly shifted towards lower energies and the energy gap seen in \NNTO~is washed out. The broad nature of the magnetic excitations is somewhat surprising, because one would naively expect that an additional structural disorder in the Na layer would have lesser impact on the spin dynamics considering the quasi-2D nature of magnetic interactions. It is thus reasonable to infer that the additional Na-content induces a charge disorder inside the Ni/Te magnetic honeycomb layer that strongly damps the magnetic excitations. Based on the macroscopic magnetic measurements and bond valence sum analysis it has been determined that the parent compound \NNTO~presents the charge arrangement Na$^{1+}_2$Ni$^{2+}_2$Te$^{6+}$O$^{2-}_6$, with the Ni in the electron configuration $e_g^2t_{2g}^6$ and $S$ = 1. The effective magnetic moments inferred from Curie–Weiss analyses and the refined static magnetic moments were found to be nearly the same for the parent and Na-doped samples. This leads us to believe that the additional charge in Na layer is satisfied by an induced mixed valence on Te site (Te$^{6+}$ and Te$^{4+}$) while Ni ions remain bivalent. This is further supported by the observation of similar Ni-O bond distances in the two Ni based compounds, while the average Te-O bond length is larger in \NxNTO~ (2.001(4) \AA) than in \NNTO~ (1.954(2) \AA). The bond distances and the valences obtained from bond valence sum calculation are given in the supplementary material.\cite{SM}~This sort of disorder is expected to primarily impact the second or third NN couplings $J_2$ and $J_3$, which are mediated by O-Te-O bridges.

The broad features in magnetic excitation spectrum presents limitations to data modeling. As a result we focus on providing a minimal quantitative model that describes the excitations using the model solutions obtained for the undopped sample. To account for the broadening of the spin-excitations the calculated spectrum was convoluted with a Gaussian function with the width of 3 meV, that is approximately six time larger than the instrumental resolution. A good description of the data can be obtained with either the \textit{XXZ} or the \textit{K-H} model, by simply scaling down to about 70\% the values of $J_2$ and $J_3$ exchange interaction obtained for the parent compound. This finding confirms the hypothesis that there is an induced charge disorder on the Te site. The calculated spin-wave spectrum using the \textit{XXZ} model (defined by Eq.~\ref{eq_2}) with rescaled $J_2$=0.056 meV and $J_3$=1.1 meV, is shown in Fig.~\ref{insNix}(b). A similar match to the experimental data is obtained from using the adapted \textit{K-H} model ``B'', where the new exchange interactions become: $J_2$=0.12 meV and $J_3$=0.92 meV. We remind the reader that solution ``B'' (with $K\approx$ 0 and $\Gamma \approx \Gamma^{\prime}$) is equivalent to the \textit{XXZ} model, but it does not include a SIA contribution. Selecting between the two \textit{XXZ} -type solutions comes to the comparison of the optimized magnetic structures using model parameters with the structure determined from the diffraction study. In that regard, the \textit{XXZ} model ``B'' seems to be better suited for \NxNTO~because it accurately predicts the canting of magnetic moments away from the $c$-axis. It thus appears that manipulation of the Na content can be an efficient way to control both the Kitaev interactions as well as the easy-axis anisotropy in these materials.

\section{Summary}

In this study we have evaluated the static order and spin dynamics of \NCTO~ and \NNTO~ honeycomb compounds. In addition, we investigated the effect of Na-doping on the magnetic behavior of the Ni-based material. Our neutron diffraction data confirmed that \NCTO~ orders magnetically with a propagation vector $\mathbf{k}$ = ($\frac{1}{2}$,~0,~0). We showed that in addition to the predominant in-plane magnetic moment component forming the zigzag-type structure, there is an  additional out-of-plane ordered component leading to a slightly canted structure. The magnetic moments are orthogonal to the propagation vector and the canting angle is estimated to be approximately 14 degrees away from the horizontal plane. More surprising results were obtained for the \NNTO~system were the magnetic order is found to be defined by the propagation vector $\mathbf{k}$ = ($\frac{1}{2}$,~0,~$\frac{1}{2}$), which is different from that reported previously. Refinements of crystal structure of our sample revealed a more ordered distribution of Na atoms, which are likely responsible for mediating competing out-of-plane magnetic exchange interactions. We also determined that magnetic order is sensitive to the Na content and that the \NxNTO~ compound orders with a wave-vector $\mathbf{k}$ = ($\frac{1}{2}$,~0,~0). In addition to the change in stacking sequence of adjacent honeycomb layers, the two Ni-compounds also present different moment orientations. In \NNTO~ the magnetic moments are aligned parallel to the $c$-axis, while in \NxNTO~ they are canted away from the $c$-axis. In both materials the moments form ferromagnetic zigzag chains coupled antiferromagnetically. The refined static magnetic moment was found to not depend much on the Na content, suggesting that Ni ions remain bivalent while the overall charge balance is stabilized by an induced mixed valence on Te site.

The spin-wave spectrum of \NCTO~was modeled using a generalized Kitaev-Heisenberg Hamiltonian. The focus in our analysis has been on addressing the conflicting reports on the sign of Kitaev coupling. To overcome the challenge in evaluating the large number of parameters involved in the \textit{K-H} model, we applied a iterative optimization procedure capable of quantifying the uncertainties of multiple model solutions over a broad parameter space. The obtained manifold of possible solutions revealed that there are two optimal regions which corresponds to either ferromagnetic ($K <$ 0) or antiferromagnetic ($K >$ 0) Kitaev parameter. Furthermore, we found that both model solutions stabilize magnetic structures with moments aligned orthogonal to the propagation vector, in agreement with the diffraction results. Our results articulate the need for single crystal data that will alow extensions of the \textit{K-H} model to more comprehensive models that take into account the anisotropy of further-nearest-neighbor couplings or multi-magnon scattering processes.

The inelastic neutron spectrum measured on \NNTO~powder exhibits an anomalous gap opening in the upper part of the spin-wave spectrum that cannot be explained using a \textit{XXZ}-Heisenberg Hamiltonian with a single-ion anisotropy. In order to reproduce that feature, we considered a $S$=1 \textit{K-H }model.~A possible solution was identified in the region of a ferromagnetic Kitaev interaction with the off-diagonal interactions ($\Gamma$, ~$\Gamma^{\prime}$) of opposite signs. This confirms the realization of $S$=1 bond-dependent Kitaev interactions that have been predicted to occur in this class of Ni$^{2+}$ materials. An interplay with a single-ion anisotropy needs to be considered to explain the spins alignment along the $c$-axis. The introduction of additional Na atoms in \NNTO~structure leads to a sizable broadening of the magnetic excitation and the disappearance of the gap feature. A plausible interpretation is that the broadening is caused by an exchange randomness due to an induced disordered valence on the Te sites, that is mostly affecting the second and third NN couplings. The valence mixing in \NxNTO~also appears to affect the effective spin-orbital coupling as well as the single-ion anisotropy of the system, allowing the moments to cant away from the $c$-axis direction. Thus, the control of Na content proves to be an efficient way to tune the bond-dependent anisotropy. A good description of the inelastic data from \NxNTO~can be obtained with an anisotropic \textit{XXZ} model, by scaling down the values of $J_2$ and $J_3$ exchange interaction obtained from \NNTO.

This study shows that the prospect of an experimental realization of Kitaev-type bond-dependent anisotropic interactions in 3$d$ electron systems remains very encouraging. Both Co$^{2+}$ with electronic $d^7$, and Ni$^{2+}$ in a $d^8$ configuration require Kitaev-Heisenberg Hamiltonian models to explain their intricate spin-dynamic spectra. More experimental studies involving single crystal sample are definitely interesting to pursue in the future.

\begin{acknowledgments}
The authors gratefully acknowledge technical assistance from M. K. Graves-Brook during HYSPEC experiments. This research used resources at the High Flux Isotope Reactor and Spallation Neutron Source, DOE Office of Science User Facilities operated by the Oak Ridge National Laboratory. Q.C. and H.D.Z. were supported by the National Science Foundation, Division of Materials Research, under Awards No. DMR-2003117.

\end{acknowledgments}

\end{document}